\documentclass[12pt]{article}
\usepackage{latexsym}
\usepackage{amsfonts}
\usepackage{amssymb}

\begin{document}
\newcommand{\proof}{{\em Proof. }}
\newcommand{\al}{\alpha}
\newcommand{\bet}{\beta}
\newcommand{\ga}{\gamma}
\newcommand{\de}{\delta}
\newcommand{\la}{\lambda}
\newcommand{\si}{\sigma}
\newcommand{\dt}{\delta^{(3)}}
\newcommand{\ed}{\hfill Q.E.D.}
\newcommand{\et}{\label}
\newcommand{\rf}[1]{(\ref{#1})}
\newcommand{\re}{\ref}
\newcommand{\be}{\begin{equation}}
\newcommand{\ee}{\end{equation}}
\newcommand{\bea}{\begin{eqnarray}}
\newcommand{\eea}{\end{eqnarray}}
\newcommand{\ra}{\rightarrow}
\newcommand{\rotf}{{\rm curl}_{\varphi}}
\newcommand{\rotfi}{{\rm curl}_{\varphi}^{-1}}
\newcommand{\di}{{\rm div}}
\newcommand{\grad}{{\rm grad}}
\newcommand{\ph}{\varphi}
\newcommand{\p}{\partial}
\newcommand{\ad}{{\dot a}}
\newcommand{\rb}{{\bf R}}
\newcommand{\bb}{{\bf b}}
\newcommand{\lb}{{\bf l}}
\newcommand{\qb}{{\bf Q}}
\newcommand{\xb}{{\bf X}}
\newcommand{\yb}{{\bf Y}}
\newcommand{\bc}{{\cal B}}
\newcommand{\dc}{{\cal D}}
\newcommand{\fc}{{\cal F}}
\newcommand{\ic}{{\cal I}}
\newcommand{\jc}{{\cal J}}
\newcommand{\bn}[1]{\bc_{(#1)}}
\newcommand{\dn}[1]{\dc_{(#1)}}
\newcommand{\bg}[1]{\bc^{(#1)}}
\newcommand{\dg}[1]{\dc^{(#1)}}
\newcommand{\eps}{\epsilon}
\newcommand{\pr}{{}^{\prime}}
\newcommand{\ip}{\rfloor}
\newcommand{\ba}{\begin{array}}
\newcommand{\ea}{\end{array}}
\newcommand{\x}[1]{\xb_{(#1)}}
\newcommand{\y}[1]{\yb_{(#1)}}
\newcommand{\ha}{{}^{\#}}
\newcommand{\spa}{{\rm span}}
\title{\bf {\huge A geometric analysis of \\
the Maxwell field in a vicinity \\of a multipole particle\\
and new special functions}}
\author{Jerzy Kijowski${}^a$ and Piotr Podle\'s${}^b$\\
${}^a$ Center for Theoretical Physics, Polish Academy of
Sciences,\\ Aleja Lotnik\'ow 32/46, 02--668 Warszawa, Poland\\
${}^b$ Department of Mathematical Methods in Physics,\\ Faculty of
Physics,  Warsaw University,\\ Ho\.za 74, 00--682 Warszawa,
Poland}
\date{}
\maketitle
\begin{abstract}
A method of solving Maxwell equations in a vicinity of a multipole
particle (moving along an arbitrary trajectory) is proposed. The
method is based on a geometric construction of a
trajectory-adapted coordinate system, which simplifies
considerably the equations. The solution is given in terms of a
series, where a new family of special functions arises in a
natural way. Singular behaviour of the field near to the particle
may be analyzed this way up to an arbitrary order. Application to
the self--interaction problems in classical electrodynamics is
discussed.
\end{abstract}

\section{Introduction}

Classical approach to the problem of motion in classical
electrodynamics is due to Dirac (cf. \cite{Dirac}, \cite{Haag}).
In this approach, equations of motion of charged point particles,
interacting {\em via} Maxwell field, are derived from field
equations by the following procedure: a generic state of the
composed ``particles + field'' system is treated as a deformation
of the ``ground state'' of the field, uniquely determined by the
positions of the particles. The ground state is defined {\em via}
a (non-local in time) decomposition of the actual field into the
{\em retarded} (or {\em advanced}) field and the remaining
``radiation field''. Unfortunately, to decide what is {\em the
retarded} or {\em the advanced} field, the entire trajectory of
the particle must be known in advance and, whence, the causality
of the theory is violated.

Recently, it was shown that such a non-causal procedure may be
avoided. In this approach, equations of motion are a simple
consequence of the conservation laws imposed on an (appropriately
defined, ``already renormalized'') total four-momentum of the
``particles + field'' system (see e.g.~\cite{K}, \cite{GKZ}, \cite
{KP}). Here, the ``ground state'' of the system may be defined
{\em via} a conditional minimization of the energy, with the
positions and the charges of the particles being fixed.

Mathematically, this leads to a simple (elliptic) variational
problem for the behaviour of the field in a topologically
non-trivial region of ${\mathbb R}^3$ (exterior of the particles),
where the charges of the particles provide the necessary boundary
conditions. Such a reduction of {\em electrodynamics} to {\em
electrostatics} provides basic information about the behaviour of
the field in the vicinity of the particle (e.~g.~the leading
$r^{-2}$ term for the monopole particle or the $r^{-3}$ behaviour
for the dipole) but fails to capture its more subtle features,
which are necessary to describe such phenomena like polarization,
which is necessary for the analysis of the dynamical stability of
the classical electrodynamics (see e.~g.~discussion of these
issues in \cite{stability} and the references therein).

A substantial improvement of description of the field singularity
near the trajectory is obtained if we reduce the dynamics with
respect to a one-parameter group of boosts (see \cite {KP}),
instead of the one-parameter group of time translations,
corresponding to the standard, flat ``(3+1)-decomposition'' of the
Minkowski spacetime. As a result, we obtain the ``electrostatics
in a uniformly accelerated reference frame'' as a proper tool to
construct the ground state of the field. Here, field equations are
again elliptic, but the flat Eucledean geometry is replaced by the
Lobaczewski geometry. Within this framework, the Born solution
describing Maxwell field attached to a uniformly accelerated
particle (see e.~g.~\cite{Th}) arises as a fundamental solution of
the Laplace-Beltrami operator $\Delta_L$ in the Lobaczewski space:
$\Delta_L \psi = \delta_0$, where the right hand side is the Dirac
delta distribution (a nice analysis of this structure was proposed
by Turakulov in \cite{Tu}).

In the present paper we show how to use this (naturally arising)
Lo\-ba\-czewski three-geometry to the analysis of the full
dynamical problem: analysis of the Maxwell field generated by a
(moving along an arbitrary trajectory) multipole particle. Our
method consists in splitting the portion of spacetime in the
vicinity of a generic trajectory of a particle into a collection
of 3-spaces orthogonal to the trajectory. Each of these spaces
carries a natural Lobaczewski geometry, uniquely implied by the
instantaneous acceleration of the particle. As a result of this
construction, the flat spacetime geometry is replaced by a curved
(general relativistic) geometry, whose parameters are uniquely
defined by the parameters of the trajectory in question. We show
that the Maxwell equations may be rewritten in terms of this
geometry. This method provides a new, powerful tool, which enables
us to solve Maxwell equations by a successive approximation
method.

The results presented here give us an important improvement and
generalization of earlier methods (see \cite{KK2} or \cite{KK}),
based on the use of the Fermi coordinates, defined by the Fermi
tetrad attached to the trajectory. The geometric construction
proposed in the present paper will be referred to as a {\em
modified Fermi system}.

Naively speaking, the standard (3+1)-approach consists in
approximating the generic trajectory by a straight line, matching
only the instantaneous velocity of the particle. In the approach
based on Fermi coordinates, we use much better approximation given
by hyperbolic (uniformly accelerated) trajectories, which match
not only the velocity but also the acceleration of the particle.
The value of acceleration is encoded in the parameter of the
Lobaczewski 3-space arising in this construction. In our new
approach (which we refer to as the ``modified Fermi''), proposed
in the present paper, we use still better approximation:
instantaneous acceleration $a(t)$ of the trajectory implies the
parameter of each of the (Lobaczewski) 3-space folia $\{ t =
$const.$\, \}$ of the constructed geometry; there is, however, a
non vanishing  {\em shift vector} whose value encodes the
derivatives ${\dot a}^j(t)$.

The basic technical tool of our method consists in deriving an
explicit formula for the operator inverse to the Lobaczewski curl
operator (which we denote by ``${\rm curl}_\varphi$'') on a
certain space of divergence-free vector fields. The inversion
requires a successive solution of Legendre differential equation
with non trivial right hand side. This, in principle, could lead
very quickly to non-elementary functions (see formula (\ref{23})
in Section \ref{XandY}). To our great surprise, we were able to
express the results (up to the order ``+1'' in the radius $r$) in
terms of rational functions of two universal combinations of the
radial coordinate, namely $u=ar/2$ and $R=\log(\frac{1-u}{1+u})$,
where $a$ is the (scalar) acceleration of the particle (see
Section \ref{Application}). This result (together with the
extremely simple and natural physical context) leads us to a
conjecture that the functions obtained this way form a new family
of special functions, which we discuss at the end of the paper.

Physically, our results form a basis for the theory of motion of a
polarizable particle: a particle which carries a dynamical
electric or magnetic moment, i.~e.~a moment which is not frozen
within the particle but depends dynamically upon the surrounding
electromagnetic field. There is a serious hope that such a theory
would be free of the standard non-stability problems accompanying
the Abraham-Dirac theory (see \cite{stability} and the discussion
therein).

The paper is organized as follows. In Section \ref{Fermi} we
define the modified Fermi system. In Section 3 we derive Maxwell
equations in the modified Fermi system and interpret them in terms
of the Lobaczewski geometry. In Section 4 we define basic fields
$\xb$ and $\yb$ and show that the operations $\di$, $\rotf$ and
$\#$ act on $\xb$, $\yb$ in a simple way. These results give a
generalization of the construction proposed in \cite{KK}. In
Section 5 we solve Maxwell equations in a vicinity of the
multipole particle, in terms of a series consisting of the $\xb$
and $\yb$ fields. The procedure is analogous to that used in
\cite{KK}. In case of the monopole and the dipole particle, the
resulting series up to the order $r^1$ is explicitely given in
Section 6, where $r$ is the radial coordinate in the modified
Fermi system. In Section 7 we propose a conjecture concerning the
special functions arising naturally in our construction. Several
specific computations and proofs are shifted to the Appendices
A-D. The results of Sections 6 and 7 were obtained partially with
help of the symbolic computations program MAPLE $8$.

Thorough the paper we use greek indices (running from $0$ to $3$)
to label space-time coordinates, and latin indices (runing from
$1$ to $3$) to label space coordinates. We always use Einstein
convention: summation over repeated indices (in both space-time
and space cases). The components of an $n$-tensor ${\bf T}$ are
denoted by $T_{i_1i_2\dots i_n}$.

\section{The Fermi--propagated and the modified \\
Fermi--propagated systems}\label{Fermi}

In this section we recall properties of the Fermi--propagated
system of coordinates, in which the particle ``remains at rest''
at each instant of time. Next we modify that system in a way which
is similar to transition from cartesian to bispherical (bipolar)
coordinates in ${\mathbb R}^3$, with the particle's position being
one of its centers (poles).

Let $y^{\lambda}$, $\lambda=0,1,2,3$, denote the (Minkowski)
spacetime coordinates in a fixed inertial ("laboratory") system,
corresponding to the metric tensor $\eta={\rm diag}(-,+,+,+)$.
Consider an arbitrary particle's trajectory
$q^{\lambda}(t)=(t,q^k(t))$. Let $\tau=\tau(t)$ denote a
particle's proper time along the trajectory. The normalized
four-velocity is given by ${\bf u}=dq/d\tau$ and the particle's
acceleration by ${\bf a}=d{\bf u}/d\tau=d^2q/d\tau^2$. We define
(see e.g. \cite{KP}) the rest--frame space $\Sigma_{\tau (t)}$ as
the hyperplane orthogonal to the trajectory (i.e.~to the vector
${\bf e}_{(0)}:={\bf u}$) at the point $q(t)$. We choose an
orthonormal basis ${\bf e}_{(l)}$, $l=1,2,3$, in $\Sigma_{\tau}$,
such that ${\bf e}_{(\mu)}$ are positively oriented. Thus $({\bf
e}_{(\alpha)}|{\bf e}_{(\beta)})=\eta_{\alpha\beta}$. Since $({\bf
u}|{\bf a})=0$, one gets ${\bf a}=a^l{\bf e}_{(l)}$ for some
$a^l$.

At each instant of time $\tau$, the above system may be subject to
the $SO(3)$ group of rotations of the {\em Dreibein} $({\bf
e}_{(l)}(\tau))$, playing role of the group of gauge
transformations. The Fermi-propagated system (see e.g.~\cite{KP})
is defined by the following condition:
\[
  d{\bf e}_{(l)}/d\tau=a^l{\bf u} \ .
\]
Once the above condition is satisfied, there remains a single,
global (time-independent) $SO(3)$ gauge transformation,
corresponding to an arbitrary choice of initial conditions: ${\bf
e}_{(l)}(\tau_0)$. Moreover, we obtain in this case:
\[
   d{\bf u}/d\tau = a^l {\bf e}_{(l)} \ ,
\]
which enables us to interpret $a^l$, $l = 1,2,3;$ (with $a^0 = 0$)
as components of the acceleration vector. Next, we define in a
neighbourhood of the trajectory the Fermi-propagated (local)
coordinates $\xi^{\mu}:=(\xi^0,\xi^l)$ putting $\xi^0:=\tau$ and
\be
 y^{\lambda}=q^{\lambda}(\tau)+\xi^l e^{\lambda}_{(l)}(\tau). \et{1}
\ee
Hence, $\xi^l$ are cartesian coordinates on $\Sigma_{\tau}$,
related to ${\bf e}_{(l)}$ and centered at the particle's position
($\xi^l=0$ for $y=q(t)$). The Minkowski metric tensor $\gamma$ in
that system is given by
\be
 \gamma_{kl}=\delta_{kl},\quad \gamma_{0l}=0, \quad
 \gamma_{00}=-N^2, \et{2}
\ee
where $k,l=1,2,3$, $N=1+a_l\xi^l$ (see e.g. \cite{KP}, p. 373, but
with $\xi$ denoted there by $x$).

Now, we change variables on each $\Sigma_\tau$ and define the
``modified Fermi-propagated'' system $x^{\mu}=(x^0,x^l)$ by
putting $x^0:=\tau=\xi^0$ and:
\be
   \et{mF}
  x^l=\frac{\xi^l+\frac12 a^l\rho^2}{1+a_i\xi^i
  +\frac14 a^2\rho^2}, \et{2a}
\ee
where $a=(\gamma_{kl}a^ka^l)^{1/2}$,
$\rho=(\gamma_{kl}\xi^k\xi^l)^{1/2}$. It is easy to check that the
inverse transformation is given by the following formula:
\be
 \xi^l=\frac{x^l-\frac12 a^l r^2}{1-a_i x^i+\frac14 a^2 r^2}, \et{2b}
\ee
where $r=(\gamma_{kl}x^kx^l)^{1/2}$ ($a_i=a^i$ is the same as
before because the particle is situated at $x=\xi=0$, where
$\partial x^k/\partial\xi^l=\delta^k_l$ and \rf2 holds).

A simple computation shows that the flat (Minkowski) metric tensor
has now the following components:
\be
  \et{metryka}
 g_{kl}=\frac{N^2}{\ph^2}\delta_{kl},\quad g_{0l}=
 \frac{\p\xi^k}{\p\tau}\frac{\p\xi^k}{\p x^l}, \quad g_{00}
 =-N^2+\frac{\p\xi^k}{\p\tau}\frac{\p\xi^k}{\p\tau},
\ee
with
\be
  \ph=1-(ar/2)^2 \et{phi}
\ee
Of course, we have: $g^{00}= g(d\tau\ ; d\tau )=-N^{-2}$, and
thus the lapse function $N$ is the same as for the previous Fermi
system, because the time variable has not been changed (cf.
\cite{KK}). In terms of the new coordinates, its value may be
written as follows: $N=\ph/M$, where we denote
$M=1-a_ix^i+\frac14a^2 r^2=(1+a_k\xi^k+\frac14a^2\rho^2)^{-1}$.
 Moreover, $\sqrt{\det
g_{kl}}=(N/\ph)^3=M^{-3}$. Denoting by $\tilde g$ the
3-dimensional inverse of $(g_{kl})$, ${k,l=1,2,3}$, the shift
vector may be calculated  as follows:(cf. \cite{KK})
\be
  \et{shift}
 N^k={\tilde g}^{kl}g_{0l}=\frac{\p x^k}{\p\xi^s}\frac{\p\xi^s}{\p\tau}
 =x^k \dot a_j x^j-\frac12r^2\dot a^k,
\ee
where ``dot'' denotes the derivative w.r.t. $\tau$, e.g. $\dot
a_j=da_j/d\tau$.

To illustrate the geometric structure of the above coordinate
system, let us apply to the above picture such a rotation ${\cal
O}$, which positions the third axis ${\bf e}_{(3)}$ in the
direction of the acceleration ${\bf a}$, i.~e.~such that we have:
${\bf a}=a\, {\bf e}_{(3)}(\tau)$. Apply now rotation ${\cal O}$
to coordinates $(x^1,x^2,x^3)$ and denote by $(z^1,z^2,z^3)$ new
coordinates on $\Sigma_{\tau}$, obtained {\em via} such a
rotation. Next, using $z^k$ as Cartesian coordinates, construct
the corresponding spherical coordinates $(r,\eta,\phi)$. Finally,
define the variable $\mu$ by
\[
 r=\frac2a \exp(-\mu).
\]
Then, it is easy to show that $(\mu, \eta, \phi)$ form the
bispherical system of coordinates on the Euclidean space
$\Sigma_{\tau}$, with the particle's position being one of its
centers: $r=0$ (i.e. $\mu\ra +\infty$). The vector connecting this
center with the other center of the bispherical system is parallel
but opposite to the acceleration and its length equals $\frac 2a$.

\section{Maxwell equations}

Maxwell equations: d$f=0$ and d$*f=\jc$, with $*$ denoting the
Hodge ``star'' operator, can be written in an arbitrary system of
coordinates, in an arbitrarily curved spacetime, as follows (cf.
e.g. \cite{KK}):
\be
 \p_{\,[\gamma} f_{\mu\nu]}=0, \et{3}
\ee
\be
 \p_\nu\fc^{\mu\nu}=\jc^{\mu}, \et{4}
\ee
\be
  \fc^{\mu\nu}=\sqrt{-\det g}\ g^{\mu\alpha}g^{\nu\beta}
  f_{\alpha\beta} \ , \et{4,5}
\ee
where the brackets ``$[ \ ]$'' denote the complete
anti-symmetrization. Moreover, we denote: $\jc^{\mu}=\sqrt{-\det
g}J^{\mu}$, where $J^{\mu}$ is the four--current vector and
$\jc^{\mu}$ is the current density (an ``odd three-form'', see
e.~g.~\cite{springer}), satisfying the continuity equation
$\p_{\mu}\jc^{\mu}=0$. The two vector--densities:  $\dc$ (electric
induction) and  $\bc$ (magnetic induction), are defined (cf.
\cite{KK}) as the following components of these tensors: $\frac 12
f_{kl} \eps^{klm}= \bc^m$, $\fc^{0k}=\dc^k$, where $\eps^{klm}$ is
the standard Levi--Civitta symbol (totally antisymmetric and
normalized: $\eps^{123}=1$). Due to \rf{metryka}, the covariant
components of the magnetic field, calculated in our modified-Fermi
coordinates \rf{mF}, are equal to (cf.~\cite{KK}):
$B_m=g_{mk}B^k=g_{mk}(\det g_{kl})^{-1/2}\bc^k=(\ph/N)\bc^m$.
Hence, we have: $NB_m=\ph\bc^m$ and, similarly, $ND_m=\ph\dc^m$.
These formulae show that a substantial simplification of the field
equations is obtained, if we use a fictitious, flat metric
$\delta_{kl}$ to identify ``upper and lower indices'', i.~e.~to
fix an isomorphism between vectors and covectors. From now on, we
strictly observe this convention. Consequently, we can rewrite
Maxwell equations \rf3--\rf{4,5} (cf. (9)-(12) of \cite{KK}) as
equations for the two quantities $(\dc,\bc)$, treated as two
vectors in this fictitious flat geometry:
\be
 \p_k\dc^k=\jc^0, \et{5}
\ee
\be
 \p_k\bc^k=0, \et{6}
\ee
\be
 \dot\dc^k-\p_l(N^l\dc^k-N^k\dc^l)=
 \eps^{kil}\p_i(\ph\, \delta_{lj}\bc^j)-\jc^k, \et{7}
\ee
\be
 \dot\bc^k-\p_l(N^l\bc^k-N^k\bc^l)=
 -\eps^{kil}\p_i(\ph\, \delta_{lj}\dc^j), \et{8}
\ee
where the vector $N^k$ is uniquely implied by the derivative of
the trajectory up to the third order, according to formula
\rf{shift}, whereas the ``conformal factor'' $\ph$ is implied by
the second derivative, according to formula \rf{phi}. We introduce
the following short-hand notation for the differential operators
appearing here:
\be
 {\bf W}^{\#}=\dot {\bf W}+\tilde {\bf W}, \et{12a}
\ee
where
\be
 \tilde W^k=\p_l(N^kW^l-N^lW^k) \ , \et{12b}
\ee
and
\[
  \rotf {\bf W} := \mbox{\rm curl} (\varphi {\bf W}) \ .
\]
This enables us to rewrite Maxwell equations as follows:
\be
 \di\dc=\jc^0, \et{9}
\ee
\be
 \di\bc=0, \et{10}
\ee
\be
 \dc^{\#}=\rotf\bc-\jc, \et{11}
\ee
\be
 \bc^{\#}=-\rotf\dc \et{12} \ .
\ee

Now, we are going to show that the differential operator
``$\rotf$'' can be nicely interpreted as a genuine curl operator
in the Lobaczewski geometry. For this purpose consider the
covectors which are subject to differentiation in the definition
of $\rotf$, see formulae \rf{7}, \rf{8}:
\[
  d_l := \ph\, \delta_{lj}\dc^j \ ,
   \quad
   b_l := \ph\, \delta_{lj}\bc^j \ .
\]
We want to find an appropriate (yet another) 3-metric $\mu$ on
each leaf $\Sigma_\tau$, such that the above covectors
(differential one-forms) are related with the vector densities
$\dc$ and $\bc$ (differential three-forms) {\em via} the standard
Hodge formula:
\be
  \et{constitutive}
  \dc^k = \sqrt{\mu} \, \mu^{kl} d_l \ ,
   \quad
   \bc^k = \sqrt{\mu} \, \mu^{kl} b_l \ .
\ee
It is easily seen that $\mu$ must be conformally similar to the
flat metric: $\mu_{kl}= \lambda \delta_{kl}$. Taken into account
that $\sqrt{\mu} \, \mu^{kl} = \lambda^{\frac 12} \delta^{kl}$, we
conclude that to fulfil \rf{constitutive} we must take $\ph
\lambda^{\frac 12} = 1$, i.~e.~$\lambda = \ph^{-2}$. It is a
matter of easy calculations to check that, indeed, the resulting
metric tensor:
\be
  \mu_{kl}:=\ph^{-2} \delta_{kl} = \frac 1{1-(ar/2)^2}\ \delta_{kl} \ ,
\ee
defined on the ball $K:= \{ {\vec x} \in {\mathbb R}^3 : r= \|
\vec x \| < \frac 2a \}$ is equal to the Lobaczewski metric with
constant negative curvature $R = - 6 a^2$.

In the particular case of a uniformly accelerated trajectory,
i.~e.~when ${\dot a} = 0$, the shift vector vanishes: $N=0$. As
noticed by several authors, the ``electrostatics'' (with respect
to such a uniformly accelerated reference system!) implies:
$\rotf\dc = 0$, equivalent to $d = \grad \ \psi$. Then, Gauss
equation \rf{9} reduces such a theory to the standard ``potential
theory'' in the Lobaczewski space: $\di \ \grad \ \psi = \Delta_L
\psi = \jc^0$. The fundamental solution of this equation,
corresponding to the point particle: $\jc^0 = \delta_0$
(Dirac-delta-like charge density -- cf.~\cite{Tu}), describes the
electromagnetic field accompanying the uniformly accelerated
particle and is equal to the Born solution (cf.~\cite{Born} or
\cite{Th})

\section{The calculus of $\xb$ and $\yb$ fields}
\label{XandY}

In this section we define elementary fields $\xb$ and $\yb$, which
enable us to reduce the asymptotic analysis of Maxwell equations
in a neighbourhood of a moving, polarized particle, to a
relatively simple algebra of these fields.

Given a pair $(n,m)$ of natural numbers such that $n=1,2,\dots$,
$m=0,1,2,\dots$, consider the vector space $L_{n,m}$ of Laurent
series of a single variable $r$, of the form
\[
 \sum_{k=0}^{\infty} c_k r^{-n+2m+2k},
\]
convergent in an annular neighbourhood $0 < r < \epsilon$, where
the coefficients $c_k$ are real. Observe that we have $L_{n,m+1}
\subset L_{n,m}$. By the order of the series we understand the
order of its first non-vanishing term (i.~e.~$f$ is of order $l$
if it behaves at $r=0$ like $r^{l}$).

 For any pair $(f,\qb)$, where $f\in
L_{n0}$, and $\qb= (Q_{i_1\cdots i_n})$ is a completely symmetric,
traceless tensor of rank $n$ (i.~e.~satisfying: $\delta^{i_1
i_2}Q_{i_1 i_2 \cdots i_n} = 0$), consider the following vector
fields:
\be
 X^k_{(n)}(f,\qb)=[(n+1)f-rf_{,r}]\frac1{r^{n+3}}
 x^kQ_{i_1\cdots i_n}x^{i_1}\cdots x^{i_n} +
 f_{,r}\frac1{r^n}Q^k_{i_2\cdots i_n}x^{i_2}\cdots x^{i_n}, \et{13}
\ee
\be
 Y^k_{(n)}(f,\qb)=\frac{f}{\ph} \frac{1}{r^{n+1}}
 \eps^{kjl}x_jQ_{li_2\cdots i_n}x^{i_2}\cdots x^{i_n} \ . \et{14}
\ee
For an exceptional case $n=0$, every $\qb$ is a scalar and,
therefore, the field $Y_{(0)}$ is not defined. Moreover, we assume
in this case that $f$ is constant.

This is an appropriate generalization of the fields $X$ and $Y$,
which were defined in \cite{KK} for purposes of the analysis of
the Maxwell equations in terms of the standard Fermi system. As
will be seen in the sequel, this generalization enables us to
describe the asymptotic behaviour of the Maxwell field in a
neighbourhood of a polarized, point particle in a  much efficient
way.

It can be easily shown (cf. Appendix A) that, outside of the
center $r=0$, the fields $X$ and $Y$ are divergence-free:
\be
 \di \xb_{(n)}(f,\qb)=\di \yb_{(n)}(f,\qb)=0, \et{14b}
\ee
and the operator $\rotf$ acts in a simple way on these fields:
\be
 \rotf \yb_{(n)}(f,\qb)=-\xb_{(n)}(f,\qb)\ , \et{19}
\ee
\be
 \rotf \xb_{(n)}(f,\qb)=-a^2 \yb_{(n)}(h_{(n)}(f),\qb), \et{20}
\ee
where
\be
  h_{(n)}(f)=-\frac{\ph^2}{a^2}f_{,rr}+\frac{\ph r}{2}f_{,r}
  +\frac{n(n+1)\ph^2}{a^2r^2}f \ . \et{16}
\ee
(The full action of these operators understood in the sense of
distributions contains also Dirac-$\delta$-like terms at $r=0$,
cf. Appendix \ref{dystrybucje}). A much simpler form of the
operator $h_{(n)}$ is obtained if we introduce the following new
variables: $u=ar/2$,
$z=\frac12(u^{-1}+u)=\frac12(\frac2{ar}+\frac{ar}2)=\cosh\mu$.
Then $z\geq1$ and assuming $u<1$ (i.~e.~near to the particle) we
have: $r=2u/a=\frac2a(z-\sqrt{z^2-1})$. Now, $h_{(n)}$ takes the
form:
\be
 h_{(n)}[g(z)]=(z^2-1)\{(1-z^2)\frac{d^2g}{dz^2}-2z\frac{dg}{dz}
 +n(n+1)g\}. \et{18}
\ee
Equation $h_{(n)}[g(z)]=0$ is equivalent to the Legendre equation
 and, therefore, has two independent
solutions: the Legendre polynomial of order $n$:
$P_n(z)=\frac1{2^n n!}\frac{d^n}{dz^n}(z^2-1)^n=A_n(r)$  and the
Legendre function $\frac12
P_n(z)\log(\frac{z+1}{z-1})+v_{n-1}(z)=B_n(r)$, where $v_{n-1}$ is
a polynomial of order $n-1$, $v_{-1}=0$, $P_n$ and $v_n$ are even
(odd) for $n$ even (odd) (cf. \cite{WW}, pp. 302 and 319). Thus
$A_n(r)$ and $B_n(r)$ are solutions of equation $h_{(n)}(f)= 0$.
$A_n$ has the same parity as $n$, $B_n$ -- the opposite parity
(one can formally define them for $|u|=\frac{a|r|}{2}<1$).

It can be proved  (see Appendix \ref{seria}) that $A_n$ is a
Laurent polynomial of $r$ with the lowest power $r^{-n}$, whereas
$B_n$ is regular at $r=0$ and its Taylor series starts from
$r^{n+1}$.

{\bf Proposition.} There exists a unique operator (the index $m$
is suppressed for simplicity) $s_{(n)}: L_{nm}\ra L_{n,m+1}$,
which is inverse to $h_{(n)}: L_{n,m+1}\ra L_{nm}$.

{\em Proof.} Because solutions of the homogeneous problem are
known, equation
\be
 h_{(n)}(f)=l \et{20a}
\ee
can be solved by the ``variation of constants'' method:
\be
  f(r)=A_n(r)\int\frac{a^2l(r)B_n(r)\mbox{\rm
 d}r}{\ph^2(A_nB_n\pr-B_nA_n\pr)}-
  B_n(r)\int\frac{a^2l(r)A_n(r)\mbox{\rm
 d}r}{\ph^2(A_nB_n\pr-B_nA_n\pr)} \et{21a}
\ee
where the Wronskian $A_nB_n\pr-B_nA_n\pr$ is proportional to
$\ph^{-1}$, so the denominators under integrals are proportional
to $\ph$. Assume now that $l\in L_{nm}$. We define $s_{(n)}(l)$ as
the right hand side of \rf{21a} with the integrals performed term
by term, using the formula $\int r^{\alpha}\mbox{\rm
d}r=\frac{r^{\alpha+1}}{\alpha+1}$ ($\alpha\neq-1$). Then
$s_{(n)}$ is linear. Moreover, the Laurent series of the function
under the first integral is odd and starts at least from
$r^{1+2m}$, so the series for the integral is even and starts at
least from $r^{2+2m}$. The Laurent series under the second
integral is even and starts at least from $r^{-2n+2m}$. Its
integral is odd and starts from $r^{-2n+2m+1}$. (This proves that,
indeed, the case $\alpha =-1$ never occurs.) Thus, $s_{(n)}(l)$ is
even or odd (depending upon $n$) and starts at least from
$r^{-n+2(m+1)}$, i.~e.~we have $s_{(n)}(l)\in L_{n,m+1}$. Any
other solution $f$ of \rf{20a} differs from $s_{(n)}(l)$ by a
combination of $A_n$ and $B_n$. But the condition $f\in L_{n,m+1}$
excludes both $A_n$ (its series starts from lower power $r^{-n}$)
and $B_n$ (its parity is different from $n$). This proves the
uniqueness of $s_{(n)}$.\ed

We conclude that, restricted to the space of those
$\xb_{(n)}(f,\qb)$ and $\yb_{(n)}(f,\qb)$, for which $f\in
L_{nm}$, the operator ``$\rotf$'' has a right inverse ``$\rotfi$''
(i.e. $\rotf\rotfi={\rm id}$) given by the following formulae:
\be
 \rotfi \xb_{(n)}(f,\qb)=-\yb_{(n)}(f,\qb), \et{22}
\ee
\be
 \rotfi \yb_{(n)}(f,\qb)=-a^{-2}\xb_{(n)}(s_{(n)}(f),\qb)\ . \et{23}
\ee
Let us notice that for $f\in L_{nm}$ being of minimal order
$(-n+2m)$, one has that:
  \[ \xb_{(n)}(f,\qb) \mbox{ is at least of order } r^{-n+2m-2},\]
\[ \yb_{(n)}(f,\qb) \mbox{ is at least of order } r^{-n+2m-1}. \]
Those orders are called the generic orders. Observe that
(generically) $\rotfi$ increases the order (in $r$) by $1$.

Now, suppose that both $f$ and ${\bf Q}$ are time-dependent. It
turns out that also time-depending operator ${\#}$ (cf.
\rf{12a}--\rf{12b}) acts in a simple way on the fields $\xb$ and
$\yb$. Indeed, we have:

{\bf Theorem 1.} For $r\neq0$
\bea
 \xb_{(n)}(f,\qb)^{\#}=-\frac12 \xb_{(n+1)}(r^2f_{,r}+(n+1)rf,
\bb
\vee \qb)+\yb_{(n)}(g_1,\bb\times \qb)+\nonumber\\
+\frac{n^2-1}{2n(2n+1)} \xb_{(n-1)}(-r^2f_{,r}+nrf, \bb\ip
\qb)+\xb_{(n)}(\dot f,\qb)+\xb_{(n)}(f,\dot \qb), \et{24} \eea
\bea \yb_{(n)}(f,\qb)^{\#}= \yb_{(n+1)}(g_2, \bb\vee \qb)-
\frac1{2(n+1)}\xb_{(n)}(\frac{fr^2}{\ph},\bb\times \qb)+\nonumber\\
+\yb_{(n-1)}(g_3,\bb\ip \qb)+\yb_{(n)}(\dot f,\qb)+\yb_{(n)}(f,\dot
\qb)+
 \yb_{(n)}(\frac{fr^2}{2\ph},a\ad \qb),\et{25} \eea
where $b_k=(a_k)\dot{}\,$, $\ad=da/d\tau=a^kb_k/a$, whereas the
symmetric traceless tensors: $\bb\vee\qb$, $\bb\times\qb$ and
$\bb\ip\qb$ of ranks (respectively) $n+1$, $n$ and $n-1$, are
built from the $n$-tensor $\qb$ and the vector $\qb$ according to
the following formulae (the letters placed below dots denote
missing indices):
\begin{eqnarray}
  (\bb\vee\qb)_{i_0\cdots i_n} &=& \frac1{n+1}\sum_{s=0}^n
 b_{i_s}Q_{i_0}\!\!{}_{\ba{c} \underbrace{\cdots}_{\hat s}\ea}
 \!\!{}_{i_n}  \nonumber \\
   &-& \frac2{(n+1)(2n+1)}\sum_{k<l}\delta_{i_ki_l}b^mQ_{mi_0}\!\!{}_{\ba{c}
 \underbrace{\cdots}_{ \hat k\hat l}\ea}\!\!{}_{i_n}, \et{25a}
\end{eqnarray}
\be
  (\bb\times\qb)_{i_1\cdots i_n}=\frac1n
  \sum_{k=1}^n\eps_{i_k}{}^{lm}b_lQ_{mi_1}\!\!{}_{\ba{c}
  \underbrace{\cdots}_{\hat k} \ea}\!\!{}_{i_n}, \et{25b}
\ee
\be
  (\bb\ip\qb)_{i_2\cdots i_n}=b^mQ_{mi_2\cdots i_n}, \et{25c}
\ee
\[ g_1=\frac{\ph}{2(n+1)} [r^2 f_{,rr}+2rf_{,r}-n(n+1)f], \]
\[ g_2=-\frac12(n+3)rf-\frac12 r^2f_{,r}-\frac{a^2fr^3}{4\ph}, \]
\[ g_3=\frac{n^2-1}{n(4n+2)}\left[(n-2)rf-r^2f_{,r}-\frac{a^2fr^3}{2\ph}\right],\]
\[ \ph=1-(ar/2)^2. \]

{\bf Remark.} For the properties of $\bb\vee\qb$,
$\bb\times\qb$, $\bb\ip\qb$ (computed in a simpler situation, but
valid universally) see \cite{KK}, pp. 304-305
(Warning: Normalization of the tensors $\bb\vee\qb$ and
$\bb\times\qb$ used in \rf{25a}--\rf{25b} differs slightly from that used in
(23)-(24) of \cite{KK}). Let us notice that
for ${\bf W}$ being $\xb$ or $\yb$ field, ${\bf W}^{\#}$ is at
least of the same generic order of $r$ as ${\bf W}$. If, moreover,
$f,\qb$ do not depend upon $\tau$, the generic order of ${\bf
W}^{\#}$ increases by at least $1$ with respect to the generic
order of ${\bf W}$.

We shall use these formulae in a special case, when $f$ depends
upon time only {\em via} the acceleration $a= a(\tau)$ contained
in the combination $u=ar/2$. Assuming, therefore, that
$f(r,\tau)=f(u)$, one can rewrite \rf{24}--\rf{25} as follows:
\bea
\xb_{(n)}(f,\qb)^{\#}=-\xb_{(n+1)}\left[u^2f_{,u}+(n+1)uf,\frac1a\bb\vee
\qb\right]+\nonumber\\
\frac1{2(n+1)}\yb_{(n)}[\ph(u^2f_{,uu}+2uf_{,u}-n(n+1)f),\bb\times \qb]+\nonumber\\
\frac{n^2-1}{n(2n+1)}
\xb_{(n-1)}\left[-u^2f_{,u}+nuf,\frac1a\bb\ip \qb\right]+
\xb_{(n)}[uf_{,u},\frac{\ad}{a}\qb]+\xb_{(n)}[f,\dot \qb],
\et{25d} \label{hX(u)}
\eea
\bea \yb_{(n)}(f,\qb)^{\#}=
-\yb_{(n+1)}\left[(n+3)uf+u^2f_{,u}+\frac{2u^3f}{\ph},\frac1a
\bb\vee
\qb\right]+\nonumber\\
-\frac2{n+1}\xb_{(n)}(\frac{fu^2}{\ph},\frac1{a^2}\bb\times \qb)+\nonumber\\
\frac{n^2-1}{n(2n+1)}\yb_{(n-1)}\left[(n-2)uf-u^2f_{,u}-\frac{2u^3f}{\ph},
\frac1a\bb\ip
\qb\right]+\nonumber\\
\yb_{(n)}[uf_{,u},\frac{\ad}{a}\qb]+\yb_{(n)}[f,\dot \qb]+
 \yb_{(n)}[\frac{fu^2}{\ph},\frac{2\ad}{a}\qb].\et{25e}
 \label{hY(u)}
\eea

The proof of Theorem 1 is given in Appendix A.

\section{Expansion of the Maxwell field}

In this section we find an asymptotic solution for the Maxwell
field in a vicinity of a time-dependent, electric
$2^n-\mbox{pole}$, moving along a given trajectory.

Given a timelike trajectory, we construct its modified Fermi frame
like in Section 2 and consider Maxwell equations \rf9-\rf{12} with
electric charge:
\be
 \jc^0=c_nQ^{i_1\dots i_n}\p_{i_1}\dots\p_{i_n}\delta^{(3)} \et{28}
\ee
and the corresponding electric current
\be
 \jc^k=-c_n\dot Q^{ki_2\cdots i_n}\p_{i_2}\dots\p_{i_n}\delta^{(3)}, \et{29}
\ee
at the right-hand side. The above four-current is conserved:
$\p_{\mu}\jc^{\mu}=0$, cf. \cite{KK}. Here,
$c_n=(-1)^n4\pi/(2n-1)!!$ is the normalization factor. For $n=0$
we assume $\jc^k = 0$ and $Q={\rm const.}$ (See Appendix C for
relation of these $2^n$--poles, defined in the modified Fermi
frame, with the standard $2^n$-poles in the Fermi frame.)\\

{\bf Theorem 2.} Maxwell equations \rf9--\rf{12} with the sources
\rf{28}-\rf{29} are solved by the following (formal) series:
\be \dc=\dc_d+\dn{-n-2}+\dn{-n}+\dn{-n+2}+\dots, \et{30} \ee
\be \bc=\bn{-n-1}+\bn{-n+1}+\bn{-n+3}+\dots, \et{31} \ee
where
\be \dc_d^i=(-1)^n\frac{4\pi n}{(2n+1)!!}Q^{ii_2\dots i_n}\p_{i_2}\dots\p_{i_n}\dt,
 \et{DD}
\ee
\be
 \dn{-n-2}=\x{n}(k_nP_n(z),\qb), \et{32}
\ee
\be \bn{k+1}=\rotfi[\dn{k}\ha],\quad k=-n-2,-n,-n+2,\dots, \et{33}\ee
\be \dn{k+1}=-\rotfi[\bn{k}\ha],\quad k=-n-1,-n+1,-n+3,\dots, \et{34}\ee
i.e. $\bn{-n-1}=\rotfi[\dn{-n-2}\ha]$,
$\dn{-n}=-\rotfi[\bn{-n-1}\ha]$,
\newline $\bn{-n+1}=\rotfi[\dn{-n}\ha]$,
\dots, $k_n=(2a)^n/
(\hskip-5pt\begin{array}{c}{\scriptstyle 2n}\\[-8pt]%
{\scriptstyle n}\end{array}\hskip-5pt)$. Moreover, $\bn{l}$ and
$\dn{l}$ are at least of order $r^l$.

{\bf Remark 1.} The fields $\bc,\dc$ and the operations
$\di,\rotf,\#$ are considered here in the sense of distributions.
On the other hand, $\rotfi$ acts on the fields $\xb$ and $\yb$ via
\rf{22}--\rf{23}. One can add to \rf{9} -- \rf{12} any regular
solution $(\dc,\bc)$ of homogeneous Maxwell equations.

{\bf Remark 2.} One has $k_nP_n(z)=r^{-n}+f_n$, where
$\lim_{r\ra0} f_n(r)r^n=0$. Thus the leading term in $\dc$ equals
$\dc_d+\xb_{(n)}(r^{-n},\qb)=-\grad\psi_n$ where
\[ \psi_n=\frac{Q^{i_1\dots i_n}x_{i_1}\dots x_{i_n}}{r^{2n+1}} \]
is the standard potential for the classical $2^n$-pole \rf{28}
with $a\equiv0$, $\qb={\rm const.}$ (cf. Appendix D).

Proof of Theorem 2 is given in Appendix D.

\section{Application to the monopole and the dipole cases}\label{Application}

In this section we apply the above formulae to the case of a
monopole and a dipole particle. We find explicitly asymptotic
expansion of the Maxwell field $(\dc,\bc)$ up to the order $r^1$.

In the monopole case we start from $\dn{-2}=\x0(1,Q)$ which is of
order $r^{-2}$, in the dipole case we start from
$\dn{-3}=\x1(k_1P_1(z),\qb)=\x1({\frac1u},\frac{a}{2}\qb)+
\x1({u},\frac{a}{2}\qb)$.
The last two components are, respectively, of orders $r^{-3}$ and
$r^{-1}$. To compute the next terms we use operations $\rotfi$ and
$\#$. While $\rotfi$ generically increases the order by one,
\rf{25d}--\rf{25e} show that $\#$ applied to $\x{n}(f,Q)$ or
$\y{n}(f,Q)$ leads to terms of different orders: in generic case,
first three terms of \rf{25d} or \rf{25e} have order increased by
one, the next two terms there have order unchanged and the last
term in \rf{25e} has order increased by two. We present the
results of computation in the monopole and in the dipole cases in
terms of $u=ar/2$ and $R=\log(\frac{1-u}{1+u})=-2u+\dots$,
obtained using the symbolic calculus provided by the program MAPLE
8. The terms of order higher than $r$ (or $u$) are denoted by
$o(u)$. We collect all the terms of generic order $r^l$ from
different parts of \rf{30} or \rf{31} and denote them,
respectively, by $\dg{l}$ or $\bg{l}$. In the monopole case we
assume that the charge $Q=\rm{const.}$

\begin{center} {\bf The monopole case} \end{center}

\be \dc=\dg{-2}+\dg1+o(u), \quad \bc=\bg0+o(u), \et{39} \ee
where
\be \dg{-2}=\x0(1,Q), \quad \dg1=\x1(f_0,\frac1{a^3}Q\dot \bb),
\quad \bg0=\y1(u,\frac1a Q\bb), \et{41} \ee with
\be f_0=-\frac14(u+u^{-1})R^2-R-u=-u^3+\dots. \et{43a} \ee

\begin{center} {\bf The dipole case} \end{center}

\be \dc=-\frac{4\pi}3\qb\dt+\dg{-3}+\dg{-1}+\dg0+\dg1+o(u), \et{44} \ee
\be \bc=\bg{-2}+\bg{-1}+\bg0+\bg1+o(u), \et{45} \ee
where
\be \dg{-3}=\x1(u^{-1},\frac{a}2\qb), \et{46} \ee
\be \dg{-1}=-\x1(f_1,\frac1{2a}\ddot \qb)+\x1(u,\frac{a}2\qb), \et{47} \ee
\bea \dg0=\x2(f_2,\frac1{2a^2}(\bb\vee \qb)\, \dot{} \ )+\y1(u,\frac1{2a}
(\bb\times \qb)\, \dot{} \ )+\nonumber\\
+ \y1(\frac{u}{1-u^2},\frac1{2a}\bb\times \dot
\qb)+\x2(f_3,\frac1{2a^2}\bb\vee\dot \qb), \et{48}\eea
\bea \dg1=-\y2(\frac{u^2}{1-u^2},\frac1{3a^2}\bb\times(\bb\vee \qb))-\x3
(f_4,\frac1{2a^3}\bb\vee(\bb\vee \qb))+\nonumber\\
-\y2(u^2,\frac3{2a^2}\bb\vee(\bb\times\qb))
-\x1(f_5,\frac{\dot a}{a^2} \dot\qb)+\nonumber\\
-\x1(f_{0},\frac{\dot a}{a^3}(\dot a\qb+\frac{a}2\dot \qb))
-\x1(f_0,\frac1{a^2}(\dot a\qb+\frac{a}2\dot\qb)\, \dot{} \ )+
\x1(f_8,\frac{\dot a^2}
{2a^5}\ddot\qb)+\nonumber\\
\x1\left[f_7,\frac{\dot a}{a^3}(\frac{\ddot\qb}{2a})\, \dot{} \
+\frac{1}{a^2}   (\frac{\dot a \ddot \qb}{2a^2})\, \dot{} \
\right]
+\x1(f_6,\frac1{a^2}(\frac{\ddot\qb}{2a})\,\ddot{}\,),\et{49}\eea
\be \bg{-2}=-\y1(u^{-1},\frac{a}2\dot \qb), \et{50}\ee
\be \bg{-1}=\frac12\y2(1,\bb\vee \qb)+\x1(u,\frac1{2a}\bb\times \qb), \et{50a}\ee
\be \bg0=-\y1(u,\dot a\qb+\frac{a}{2}\dot \qb)+
\y1(f_1,(\frac{\ddot \qb}{2a})\, \dot{} \ )
+\y1(f_9,\frac{\dot a}{2a^2}\ddot\qb),\et{50b}\ee
\bea \bg1=\frac32\y2(u^2,\bb\vee \qb)-\y2(f_{10},\frac{\ad}{2a^3}(\bb\vee\qb)\,
\dot{}\ ) -\nonumber\\
\y2(f_2,[\frac1{2a^2}(\bb\vee\qb)\, \dot{} \ ]\, \dot{} \ )
 -\x1(f_0,\frac{\dot a}{2a^4}(\bb\times\qb)\, \dot{} \ )-\nonumber\\
  \x1(f_0,\frac1{a^2}[\frac1{2a}(\bb\times\qb)\, \dot{} \ ]\, \dot{} \ )
-\y2(f_{11},\frac{\dot a}{2a^3}\bb\vee\dot\qb)-\nonumber\\
\y2(f_3,(\frac1{2a^2}\bb\vee\dot\qb) \, \dot{} \ )
-\x1(f_{12},\frac{\dot a}{2a^4}\bb\times\dot\qb)- \nonumber\\
\x1(f_5,\frac1{a^2} (\frac1{2a}\bb\times\dot\qb)\, \dot{} \ )
-\y2(f_{13},\frac1{2a^2}\bb\vee\ddot\qb)+\nonumber\\
\x1(f_{14},\frac1{8a^3}\bb\times\ddot\qb), \et{50c}\eea where
$f_0$ is given by \rf{43a},
\[ f_1=-\frac14(u^{-1}+u)R^2-R+u=2u+\dots,\]
\[f_2=\left[-\frac3{16}(u^{-1}+u)^2+\frac14\right]R^2-\frac34(u^{-1}+u)R-\frac34
=u^2+\dots,\]
\[f_3=-\frac18(3u^{-2}+2+3u^2)R^2-\frac32(u^{-1}+u)R-\frac{3-5u^2+6u^4}{2(1-u^2)}=
3u^2+\dots,\]
\begin{eqnarray*}
   f_4 &=&-\frac3{32}(5u^{-3}+3u^{-1}+3u+5u^3)R^2-\frac18(15u^{-2}+14+15u^2)R \\
 &-&\frac{15u^{-1}+4u-23u^3+20u^5}{8(1-u^2)}=\frac{10}3u^3+\dots,
\end{eqnarray*}
\[f_5=-\frac{u^3}{1-u^2}=-u^3+\dots,\]
\[f_6=\frac{1}{48}(u^{-1}+u)R^4+\frac16R^3-\frac14(u^{-1}+u)R^2-R-u=-2u^3+\dots,\]
\[f_7=-\frac{1}{2u}R^2-\frac{2}{1-u^2}R-\frac{2u+4u^3}{1-u^2}=-2u^3+\dots,\]
\[f_8=\frac{1}{48}(u^{-1}+u)R^4+\frac16R^3-\frac38(u^{-1}+u)R^2-\]
\[ \frac{3-4u^2+3u^4}{2(1-u^2)^2}R-\frac{3u-3u^3+2u^5}{2(1-u^2)^2}=-2u^3+\dots,\]
\[f_9=\frac14(u^{-1}-u)R^2+\frac{1+u^2}{1-u^2}R+\frac{3u-u^3}{1-u^2}=2u+\dots,\]
\[f_{10}=\frac38(u^{-2}-u^2)R^2+\frac{3u^{-1}-2u+3u^3}{2(1-u^2)}R+\frac{3(1+u^2)}
{2(1-u^2)}=2u^2 +\dots,\]
\[f_{11}=\frac34(u^{-2}-u^2)R^2+\frac{3u^{-1}-2u+3u^3}{1-u^2}R+
\frac{3+2u^2-15u^4+6u^6}{(1-u^2)^2}
=6u^2+\dots,\]
\[f_{12}=-\frac{1}{16}(u^{-1}+u)R^2-\frac14R-\frac{u+u^3}{4(1-u^2)^2}=-u^3+\dots,\]
\[f_{13}=-\frac14(1+3u^2)R^2-\frac{u-3u^3}{1-u^2}R+\frac{5u^2-3u^4}{1-u^2}
=6u^2+\dots,\]
\[f_{14}=-\frac1u R^2-\frac4{1-u^2}R-\frac{4(u+u^3)}{1-u^2}=\frac{20}9u^5+\dots .\]

\section{Description of the functions appearing in the Maxwell
fields (a conjecture)}\label{conjecture}

The method of analyzing Maxwell equations in a neighbourhood of a
freely moving electric multipole leads, in case of a monopole and
a dipole particle, to an interesting family of relatively simple
functions of the variable $r$. These functions are obtained from
$A_n(r)=P_n(z)$ contained in \rf{32} by consecutive application of
the differential operators contained in $\#$ and the integral
operator contained in $\rotfi$. The latter is based on solving the
differential equation \rf{20a} and could {\em a priori} lead very
quickly to non-elementary functions. An unexpected result of our
analysis consists in the fact, that the functions arising here are
polynomials of the universal quantity
$R=\log(\frac{1-u}{1+u})=\log(\frac{2-ar}{2+ar})$, with
coefficients being rational combinations of $u=ar/2$ itself. This
fact leads us to a conjecture, that a new family of special
functions arises here, which makes the physical picture of the
field relatively simple.

In this Section we present a conjecture about the smallest algebra
of functions $f$ appearing in $\x{k}(f,\qb)$, $\y{k}(f,\qb)$, when
we solve Maxwell equations with multipole particles, using the
method proposed in this paper.

We set $R_s=\frac {R^s}{2^s s!}$. First we notice that only
constant functions appear in $\x0$ (remember that $\y0 = 0$).
Indeed, the leading term for the monopole particle contains only
$P_0(z)=1$ and the terms with $\x0$ produced by (\ref{hX(u)}) --
(\ref{hY(u)}) vanish.

We define $H_k$ as the linear span of functions $f$ entering
$\x{k}(f,\qb)$, $\y{k}(f,\qb)$, which appear when we consecutively
use formulae \rf{22} -- \rf{23} or (\ref{hX(u)}) -- (\ref{hY(u)}),
starting from multipole fields \rf{32}. More precisely, we take
the following

{\bf Definition.} We set $H_k$, $k=1,2,\dots$, as the smallest
vector spaces such that
\be u^{-k},u^{-k+2},\dots u^k\in H_k, \et{50d}\ee
and such that for any $f\in H_k$ one has $s_{(k)}(f)\in H_k$,
\be \frac{u^2f}{1-u^2}\in H_k,\, uf_{,u}\in H_k,\,(1-u^2)
[u^2f_{,uu}+2uf_{,u}-k(k+1)f]\in H_k, \et{58} \ee
\be u^2f_{,u}+(k+1)uf\in H_{k+1},\, u^2f_{,u}+(k+3)uf
+\frac{2u^3f}{1-u^2}\in H_{k+1}, \et{57} \ee
\be -u^2f_{,u}+kuf\in H_{k-1},\, u^2f_{,u}-(k-2)uf
+\frac{2u^3f}{1-u^2}\in H_{k-1} \quad (k\neq1). \et{59} \ee

{\bf Remark.} Due to condition \rf{50d} we have:
\[
  P_n(z)=P_n(\frac 12(u^{-1}+u))\in\spa\{u^{-n},u^{-n+2},\dots,u^n\} \ ,
\]
and, whence, functions $f$ contained in
$\dn{-n-2}=\x{n}(k_nP_n(z),\qb)$ are in $H_n$. Also for $n=0$ and
$f$ appearing in $\bn{-1}=\y1(u,\frac1a Q\bb)$ we have $f=u\in
H_1$. It remains to consider operations $\rotfi$ and $\#$ which
yield $\bn{l}$ and $\dn{l}$ with higher $l$. Due to
\rf{22}--\rf{23} and \rf{25d}--\rf{25e}, conditions
\rf{58}--\rf{59} imply that also $f$ appearing in those $\bn{l}$,
$\dn{l}$ belong to the corresponding $H_k$.

It is easy to see that conditions \rf{58}--\rf{59} can be
reformulated in a simpler, equivalent form:
\be \frac{f}{1-u^2}\in H_k,\, uf_{,u}\in H_k,\, (1-u^2)[u^2f_{,uu}
+2uf_{,u}-k(k+1)f]\in H_k,\et{61} \ee
\be \frac{uf}{1-u^2}\in H_{k+1},\, u^2f_{,u}+(k+1)uf\in H_{k+1}, \et{60} \ee
\be \frac{uf}{1-u^2}\in H_{k-1},\, u^2f_{,u}-kuf\in H_{k-1}
\quad (k\neq1). \et{62} \ee

{\bf Conjecture:} The space $H_k$ can be represented as
\be H_{k}=\bigcup_{m>k} H_{km} \quad \mbox{(increasing sequence of vector spaces)},
\et{55}
\ee
where $H_{km}$ are the direct sums:
\be H_{km}=\bigoplus_{j=0}^{\infty} H_{kjm}, \et{56} \ee

\be
  H_{k0m}=\spa\left\{\frac{u^{-k}}{(u^2-1)^m},
  \frac{u^{-k+2}}{(u^2-1)^m},\dots, \frac{u^{k+2m}}{(u^2-1)^m}\right\},
  \quad m=0,1,2,\dots, \et{52}
\ee
\be
  H_{k,2l-1,m}=\spa\left\{ \frac{R^{2l-1}u^{k+1}}{(u^2-1)^m},
  \frac{R^{2l-1}u^{k+3}}{(u^2-1)^m},\dots,
  \frac{R^{2l-1}u^{2m-k-1}}{(u^2-1)^m}\right\},\quad (m>k), \et{53}
\ee
\bea
  H_{k,2l,m}=\spa\left\{\frac{w_k^l(R,u)}{u^k(u^2-1)^m},
  \frac{w_{k-1}^l(R,u)}{u^{k-2}(u^2-1)^m},\dots,
  \frac{w_1^l(R,u)}{u^{-k+2}(u^2-1)^m}\right., \nonumber\\
  \frac{R^{2l}u^{k}}{(u^2-1)^m},
  \frac{R^{2l}u^{k+2}}{(u^2-1)^m},\dots,
  \frac{R^{2l}u^{2m-k}}{(u^2-1)^m},
  \frac{w_1^l(R,u^{-1}) u^{2m-k+2}}{(u^2-1)^m},\nonumber\\
  \left. \frac{w_2^l(R,u^{-1}) u^{2m-k+4}}{(u^2-1)^m},\dots,
  \frac{w_k^l(R,u^{-1}) u^{2m+k}}{(u^2-1)^m}\right\}, \quad (m>k),
  \et{54}
\eea
$l=1,2,\dots$, with $w_k^l$ given recursively by
$w_0^l(R,u)=R_{2l}$,
\be
  w_k^l(R,u)=\left(\sum_{j=1}^{\min(k,2l-1)} A_{kj}
  w_{k-j}^{l-[j/2]}(R,u)u^{2j-2}\right)+ (2k-3)!!R_{2l-1}u^{2k-1},
\et{51}
\ee
$k,l=1,2,\dots$ (here, we put $(-1)!!=1$, so that $(-1)!!\cdot
1=1!!$), for some real numbers $A_{kj}$, $k=1,2,\dots$,
$j=1,2,\dots,k$. Moreover, we conjecture that $A_{k1}=2k-1$,
$A_{k2}=1$, $A_{kk}=(2k-3)!!$. Unfortunately, we were not able to
find a general formula for $A_{kj}$.

{\bf Remark 1.} We know only a recursive procedure to find
$A_{kj}$: Suppose we have found all $A_{rj}$, $r<k$. Then we can
compute $w_r^l$, $r<k$ (cf. \rf{51}). We set
\[ f=\frac{w_{k-1}^l(R,u^{-1})u^{2m+k-2}}{(u^2-1)^m}, \]
i.e. $f$ is the last but one element of \rf{54}. Then we compute
$g\in H_k$ by means of the last formula of \rf{61}. Next we
decompose $g$ into the bases of $H_{k,j,m+1}$, $j\leq 2l$, where
$w_k^l$ is given by \rf{51}. This gives us an equation which may
be used to find coefficients $A_{kj}$, $j=1,2,\dots,k$. That kind
of procedure was implemented by us  for $k \le 10$, using the
symbolic calculus provided by the program MAPLE 8. In particular,
we have obtained the following results for $k \le 6$:
\be A_{11}=1, \et{63} \ee
\be A_{21}=3,\, A_{22}=1, \et{64} \ee
\be A_{31}=5,\, A_{32}=1,\,A_{33}=3, \et{65} \ee
\be A_{41}=7,\,A_{42}=1,\,A_{43}=50,\,A_{44}=15, \et{66} \ee
\be A_{51}=9,\,A_{52}=1,\,A_{53}=1273/9,\,A_{54}=485/9,\,A_{55}=105, \et{67} \ee
\be A_{61}=11,A_{62}=1,A_{63}=62564/225,A_{64}=2703/25,A_{65}=9985/3,A_{66}=945. \et{68} \ee
Using \rf{51}, one gets that
\be w_k^l(R,u)=\sum_{s=0}^{{\scriptstyle {\rm min}}(k,2l-1)}
g_{ks}(u) R_{2l-s}\,, \et{69} \ee where $g_{ks}$ are polynomials.
In particular, we have:
\be w_0^l(R,u)=R_{2l}, \et{70} \ee
\be w_1^l(R,u)=R_{2l}+uR_{2l-1}, \et{71} \ee
\be w_2^l(R,u)=3R_{2l}+(3u+u^3)R_{2l-1}+u^2R_{2l-2}, \et{72} \ee
\be w_3^l(R,u)=15R_{2l}+(15u+5u^3+3u^5)R_{2l-1}+(6u^2+3u^4)R_{2l-2}+u^3R_{2l-3}. \et{73} \ee
(Observe that, due to \rf{69}, for $l=1$ we omit the last term in
$w_2^l$ and the last two terms in $w_3^l$.)

{\bf Remark 2.} One can prove that for $A_{k1}=2k-1$ and any value
of the remaining coefficients $A_{kj}$, the elements
\rf{52}--\rf{54}, spanning each vector space $H_{kjm}$ are
linearly independent and the direct sum condition is obvious.
Moreover, using \rf{51},
 $H_{km}$ form an increasing
sequence of vector spaces and therefore $H_k$ are well defined.
The elements \rf{50d} belong to $H_{k00}\subset H_{k0m}\subset
H_k$ ($m>k$). Using once again \rf{51}, it can be also checked
that the first operations of \rf{61}--\rf{62} provide elements of
$H_k$, $H_{k+1}$ and $H_{k-1}$, respectively. It remains to find
the formula for $A_{kj}$ such that the remaining four relations of
\rf{61}--\rf{62} are fulfilled, to prove that $s_{(k)}(f)\in H_k$
for $f\in H_k$ and to show the minimality of $H_k$.

{\bf Remark 3.} Elements of $H_k$ are rational functions of $u$
and $R$. However, the generators of these spaces, proposed in formulae
\rf{52}--\rf{54} are quite complicated. But splitting the
nominators into monomials doesn't simplify the situation, because
application of $s_{(k)}$ (which must preserve $H_k$) to the
components obtained in that way leads to functions which seem to
be in general not rational in $u$ and $R$. Thus the form
\rf{52}--\rf{54} of the generators seems to be the simplest one.

\begin{appendix}

\section*{Appendices}

\section{Properties of fields $\xb$ and $\yb$.}

We assume $r\neq0$. Using \rf{13}--\rf{14}, after some
computations we get
\[ \p_lX_k=[-r^2f_{,rr}+(2n+3)rf_{,r}-(n+1)(n+3)f]\frac{x_lx_k}{r^{n+5}}H+\]
\[[(n+1)f-rf_{,r}]\frac{1}{r^{n+3}}\delta_{kl}H+n[(n+1)f-rf_{,r}]\frac{x_kQ_l}{r^{n+3}}+\]
\[ (rf_{,rr}-nf_{,r})\frac{x_lQ_k}{r^{n+2}} + \frac{f_{,r}}{r^n}(n-1)Q_{kl}, \]
\[\p_l Y_k=[\frac{rf_{,r}}{\ph}+\frac{a^2r^2f}{2\ph^2}-\frac{(n+1)f}{\ph}]\frac{x_lE_k}{r^{n+3}}+  \]
\[ \frac{f}{\ph r^{n+1}}\eps_{kl}{}^mQ_m+\frac{(n-1)f}{\ph r^{n+1}}F_{kl},\]
where $H=Q_{i_1\dots i_n}x^{i_1}x^{i_2}\dots x^{i_n}$,
$Q_k=Q_{ki_2\dots i_n}x^{i_2}\dots x^{i_n}$, $Q_{kl}=Q_{kli_3\dots
i_n}x^{i_3}\dots x^{i_n}$, $E_k=\eps_k{}^{jm}x_jQ_m$,
$F_{kl}=\eps_{k}{}^{jm}x_jQ_{ml}$. That implies \rf{14b},
\rf{19}--\rf{20},
\be x_lX^l=\frac{(n+1)f}{r^{n+1}}H,\quad x_lY^l=0 \et{A.1}\ee
and (cf. \rf{12b})
\bea \tilde X_k=\frac12[r^2f_{,rr}+rf_{,r}-(n+1)^2f]\frac{x_k(b^l x_l)H}{r^{n+3}} +
\nonumber\\
-\frac12[rf_{,r}+(n+1)f]\frac{b_k H}{r^{n+1}}-\frac12[(n-2)rf_{,r}-n(n+1)f]\frac{(b^mQ_m)x^k}{r^{n+1}} + \nonumber\\
-\frac12[rf_{,rr}+(n+2)f_{,r}]\frac{(b^lx_l)Q_k}{r^n}+
\frac{n-1}{2}\frac{f_{,r}}{r^{n-2}}b^lQ_{kl}, \et{A.2} \eea
\bea \tilde Y_k=-\frac12[\frac{rf_{,r}}{\ph}+\frac{a^2r^2f}{2\ph^2}+\frac{(n+3)f}{\ph}]
\frac{E_k(b^lx_l)}{r^{n+1}}+\nonumber\\
\frac{f}{2\ph r^{n-1}}G_k+\frac{(n-1)f}{2\ph
r^{n-1}}b^lF_{kl}+\frac{fx_k}{\ph r^{n+1}}(b^lE_l), \et{A.3}\eea
where $G_k=\eps_k{}^{lm}b_lQ_m$. Let us notice that $Q^m{}_{m}=0$
($\qb$ is traceless) implies
\[ 0=2\delta_{ms}Q^{ms}\eps^{al}{}_k x_ab_l=D_{ijs}\eps^{ijs}\eps^{al}{}_k x_ab_l, \]
where $D_{ijs}=\eps_{ij}{}^mQ_{ms}$. That and
\[ \eps_{ijs}\eps_{alk}=\de_{ia}(\de_{jl}\de_{sk}-\de_{jk}\de_{sl})+ \de_{il}(\de_{jk}\de_{sa}-\de_{ja}\de_{sk})+\de_{ik}(\de_{ja}\de_{sl}-\de_{jl}\de_{sa}) \]
give
\be \eps^{alm}x_ab_lQ_{mk}=-b^lF_{kl}+G_k. \et{A.4}\ee
Using \rf{25a}--\rf{25c} and \rf{A.4}, we obtain
\bea X_{k(n+1)}(\al,\bb\vee\qb)=[(n+2)\al-r\al_{,r}]\frac{x_k}{r^{n+4}} [(b^lx_l)H-\frac{n}{2n+1}r^2(b^mQ_m)] +\nonumber\\
\frac{\al_{,r}}{r^{n+1}}\{\frac1{n+1}[b_kH+nQ_k(b^lx_l)]-\nonumber\\
\frac2{(n+1)(2n+1)} [nx_k(b^mQ_m)+\frac{n(n-1)}{2}r^2b^mQ_{mk}]\},
\et{A.5}\eea
\be Y_{k(n)}(\bet,\bb\times\qb)=\frac{\beta}{n\ph r^{n+1}}[nb_kH-(n-1)x_k(Q^mb_m) -n(b^lx_l)Q_k+(n-1)r^2b^lQ_{kl}], \et{A.6}\ee
\be X_{k(n-1)}(\ga,\bb\ip\qb)= (n\ga-r\ga_{,r})\frac{x_k(b^mQ_m)}{r^{n+2}}+\frac{\ga_{,r}}{r^{n-1}}b^mQ_{mk} \quad (n\geq2), \et{A.7}\ee
\be Y_{k(n+1)}(\psi,\bb\vee\qb)= \frac{\psi}{\ph r^{n+2}}[\frac{1}{n+1}(\eps_k{}^{jl}x_jb_l)H+\frac{n}{n+1}E_k(b^lx_l)- \frac{n(n-1)}{(n+1)(2n+1)}r^2b^lF_{kl}], \et{A.8}\ee
\be X_{k(n)}(\kappa,\bb\times\qb)=\frac{\kappa_{,r}}{r^n}[G_k-\frac{n-1}nb^lF_{kl}] -[(n+1)\kappa-r\kappa_{,r}]\frac{x_k(b^lE_l)}{r^{n+3}}, \et{A.9} \ee
\be Y_{k(n-1)}(\rho,\bb\ip\qb)=\frac{\rho}{\ph r^n}b^lF_{kl} \quad (n\geq2). \et{A.10}\ee
Computing $\eps_l{}^{km}\eps_m{}^{ij}\eps_j{}^{ac}x_kx_iQ_ab_c$ by
means of
$\eps_{lk}{}^m\eps_{mij}=\de_{li}\de_{kj}-\de_{lj}\de_{ki}$ or
$\eps_{mi}{}^{j}\eps_{jac}=\de_{ma}\de_{ic}-\de_{mc}\de_{ia}$, one
obtains
\be (\eps_k{}^{jl}x_jb_l)H=(b^lx_l)E_k-x_k(b^lE_l)-r^2G_k. \et{A.10a}\ee

{\em Proof of Theorem 1.} Comparing \rf{A.2}--\rf{A.3} with
\rf{A.5}--\rf{A.10} and using \rf{A.10a}, we get
\be \tilde X_k=X_{k(n+1)}(\al,\bb\vee\qb)+Y_{k(n)}(\bet,\bb\times\qb)+X_{k(n-1)}(\ga,\bb\ip\qb), \et{A.11}\ee
\be \tilde Y_k=Y_{k(n+1)}(\psi,\bb\vee\qb)+X_{k(n)}(\kappa,\bb\times\qb)+Y_{k(n-1)}(\rho,\bb\ip\qb), \et{A.12}\ee
for $\al=-\frac12 [r^2f_{,r}+(n+1)rf]$, $\bet=g_1$,
$\ga=\frac{n^2-1}{2n(2n+1)}(-r^2f_{,r}+nrf)$, $\psi=g_2$,
$\kappa=-\frac1{2(n+1)}\frac{fr^2}{\ph}$, $\rho=g_3$.

Using \rf{13}--\rf{14}, we obtain that $\dot X_k$ gives the last
two terms in \rf{24} while $\dot Y_k$ gives the last three terms
in \rf{25}. That, \rf{12a} and \rf{A.11}--\rf{A.12} prove
\rf{24}--\rf{25}. \ed

\section{Proof of the properties of $A_n$ and $B_n$}
\label{seria}

The quantity
\[
z^k=\left[
\frac12\left(\frac{2}{ar}+\frac{r}{2a}\right)\right]^k=2^{k-1}a^{-k}r^{-k}+\ldots
\]
is of order $r^{-k}$. Therefore $P_n(z)$ (with the highest term of
order $z^n$) is a Laurent polynomial of order $r^{-n}$ and
analogously $v_{n-1}$ is of order $r^{-n+1}$. Moreover,
\be
  \frac12\log\frac{z+1}{z-1}=\log\frac{1+u}{1-u}=ar+\dots \et{21}
\ee
and, therefore, $B_n(r)$ can be written as a Laurent series in
$r$, starting from (at least) $r^{-n+1}$. Inserting this series
into $h_{(n)}(B)=0$, using \rf{16} and denoting the order of
$B_n(r)$ by $l\geq -n+1$, we get that vanishing of the $r^{l-2}$
term in $h_{(n)}(B)$ implies $l=-n$ (which is impossible) or
$l=n+1$. Hence, $B_n(r)$ is of order $r^{n+1}$.

\section{The transformation of $2^n$--poles ($n=0,1,2$)}

Let $\ic^{\la}$ be the current density in the Fermi frame
(coordinates $\xi$) and $\jc^{\mu}$ be the corresponding current
density in the modified Fermi frame (coordinates $x$). Then
\[
  \jc^{\mu}=\frac{\p x^{\mu}}{\p \xi^{\la}}\left| \det\left(\frac{\p\xi}{\p x}
  \right)\right| \ic^{\la}.
\]
Integrating with the test function $f(x)$, we obtain
\be S^{\mu}\equiv\int\jc^{\mu}(x)f(x)d^3x=\int \frac{\p x^{\mu}}{\p \xi^{\la}}\ic^{\la} |\det(\frac{\p\xi}{\p x})|f(x)d^3x=\int\frac{\p x^{\mu}}{\p \xi^{\la}}\ic^{\la}(\xi)\tilde f(\xi)d^3\xi, \et{B.1}\ee
where $\tilde f(\xi)=f(x(\xi))$. Differentiating \rf{2a}, one
obtains
\be \frac{\p x^{k}}{\p \xi^{l}}=(\de^k_l+a^k\xi_l)M-(\xi^k+\frac12a^k\rho^2)(a_l+\frac12a^2\xi_l)M^2, \et{B.2}\ee
\be \frac{\p x^k}{\p \xi^{0}}=\frac12\dot a^k\rho^2M-(\xi^k+\frac12a^k\rho^2)(\dot a_m\xi^m+\frac12\dot a_ma^m\rho^2)M^2, \et{B.3}\ee
where $M=(1+a_k\xi^k+\frac14a^2\rho^2)^{-1}$. Moreover, $\frac{\p
x^0}{\p \xi^l}=0$, $\frac{\p x^0}{\p \xi^0}=1$. Thus
\be S^0=\int\ic^0(\xi)\tilde f(\xi)d^3\xi, \et{B.4} \ee
\be S^k=\int L^k(\xi)\tilde f(\xi)d^3\xi, \et{B.5} \ee
where
\be L^k=\frac{\p x^k}{\p \xi^{l}}\ic^l+\frac{\p x^k}{\p \xi^{0}}\ic^0. \et{B.6} \ee

Suppose that $\ic^{\la}=\ic^{\la}_{(n)}(\xi,\qb)$ is the standard
$2^n$-pole, given by the right hand sides of the formulae
\rf{28}--\rf{29} with respect to the coordinates $\xi^{\mu}$. We
shall calculate the corresponding value of
$\jc^{\mu}=\jc^{\mu}_{(n)}$ by means of \rf{B.1} in the case of a
monopole, dipole and quadrupole, i.~e.~for $n=0,1,2$ (for $n=0$ we
assume $Q={\rm const.}$).

For $n=0$ $\ic^0_{(0)}(\xi,Q)=4\pi Q\dt$, $\ic^k_{(0)}(\xi,Q)=0$,
hence $S^0=4\pi Q\tilde f(0)=4\pi Qf(0)$, $L^k=0$, $S^k=0$,
\be \jc^{\mu}_{(0)}=\ic^{\mu}_{(0)}(x,Q),  \et{B.7}\ee
where $\ic^{\lambda}_{(n)}(x,\qb)$ is given by \rf{28}--\rf{29}
with respect to the coordinates $x^{\mu}$.

For $n=1$ $\ic^0_{(1)}(\xi,\qb)=-4\pi Q^k\p_k\dt$,
$\ic^k_{(1)}(\xi,\qb)=4\pi\dot Q^k\dt$, hence $S^0=4\pi
Q^k(\p_k\tilde f)(0)=4\pi Q^k(\p_kf)(0)$, $L^k=4\pi\dot Q^k\dt$,
$S^k=4\pi\dot Q^k\tilde f(0)=4\pi\dot Q^k f(0)$,
\be \jc^{\mu}_{(1)}=\ic^{\mu}_{(1)}(x,\qb). \et{B.8}\ee

For $n=2$ $\ic^0_{(2)}(\xi,\qb)=\frac{4\pi}{3} Q^{kl}\p_k\p_l\dt$,
$\ic^k_{(2)}(\xi,\qb)=-\frac{4\pi}{3}\dot Q^{kl}\p_l\dt$, hence
$S^0=\frac{4\pi}{3} Q^{kl}(\p_k\p_l\tilde f)(0)=\frac{4\pi}{3}
Q^{kl}(\p_k\p_lf)(0)-\frac{8\pi}{3}Q^{kl}a_l(\p_kf)(0)$,
$L^k=\linebreak -\frac{8\pi}{3}Q^{km}\dot
a_m\dt-\frac{8\pi}{3}\dot Q^{km}a_m\dt-\frac{4\pi}{3}\dot
Q^{kl}\p_l\dt$, $S^k=-\frac{8\pi}{3}(Q^{kl}a_l)\, \dot{} \
f(0)+\linebreak \frac{4\pi}{3}\dot Q^{kl}(\p_lf)(0)$,
\be \jc^{\mu}_{(2)}=\ic^{\mu}_{(2)}(x,\qb)+\ic^{\mu}_{(1)}(x,{\bf P}), \et{B.9}\ee
where the dipole charge $P^k=-\frac23 Q^{kl}a_l$. Therefore the
Fermi frame quadrupole has a dipole component in the modified
Fermi frame.

\section{Integrals, distributions and the proof of Theorem 2.}
\label{dystrybucje}

First we compute the following integral over two--sphere:
\be
S_{klm}=\int_{S^2(r)}\left(\frac{x}r\right)^{2k}\left(\frac{y}r\right)^{2l}\left(\frac{z}r\right)^{2m}d\si,
\quad k,l,m=0,1,2,\dots, \et{C.1}\ee where $r=(x^2+y^2+z^2)^{1/2}$
is the radius of the two--sphere, $d\si=\sin\theta d\theta d\phi$,
$r,\theta,\phi$ are the spherical coordinates.

{\bf Proposition 1.} One has
\[ S_{klm}=\frac{(2k-1)!!(2l-1)!!(2m-1)!!}{(2k+2l+2m+1)!!}4\pi. \]
{\em Proof.} Let $A>0$. We use
\[ \int_{\rb^3} x^{2k}e^{-Ax^2}y^{2l}e^{-Ay^2}z^{2m}e^{-Az^2}dxdydz=\int_0^{\infty}r^{2(k+l+m)}e^{-Ar^2}r^2S_{klm}dr, \]
\[ \int_{\rb} x^{2k}e^{-Ax^2}dx=(-d/dA)^k\int_{\rb} e^{-Ax^2}dx=\frac{(2k-1)!!}{2^k}\pi^{1/2}A^{-k-1/2} \]
and the analogous formulae for $y,z$ and $r$.  \ed

{\bf Corollary.}
\be \int_{S^2(r)} x_{i_1}x_{i_2}\dots x_{i_{2n}}d\si=\frac{4\pi r^{2n}}{(2n+1)!} \sum_{\la\in \Pi_{2n}} \de_{i_{\la(1)}i_{\la(2)}}\dots\de_{i_{\la(2n-1)}i_{\la(2n)}}. \et{C.2}\ee
{\em Proof.} We get nonzero results only if $i_s=1$ in $2k$ cases,
$i_s=2$ in $2l$ cases, $i_s=3$ in $2m$ cases, $k+l+m=n$. Then the
left hand side gives $S_{klm}r^{2n}$ and the sum gives the factor
$(2k-1)!!(2l-1)!!(2m-1)!!2^n n!$. Next we use \rf{C.1}. \ed

{\bf Proposition 2.} Let $Q^{i_1\dots i_n}$ be a symmetric
traceless tensor, $S^{j_1\dots j_m}$ be any tensor. We set
$Qx\cdots x=Q^{i_1\dots i_n}x_{i_1}\cdots x_{i_n}$ and similarly
for $S$. Then for $m=n$
\be \int_{S^2(r)} (Qx\cdots x)(Sx\cdots x)\, d\si=Q^{i_1\dots i_n}S_{i_1\dots i_n}\frac{4\pi(n!)^2 2^n}{(2n+1)!}r^{2n}, \et{C.3} \ee
while for $m<n$ the left hand side of \rf{C.3} equals zero.

{\em Proof.} Let $m=n$. Then the left hand side of \rf{C.3} is
equal to $Q^{i_1\dots i_n}S^{i_{n+1}\dots i_{2n}}$ multiplied by
\rf{C.2} and summed over all $i_k$. We may first sum over all
$i_k$ and then over $\la\in \Pi_{2n}$. In such a case, due to the
traceless condition for $\qb$, nonzero terms in the sum over $\la$
are obtained if for each $k=1,\dots,n$ one of elements
$\la(2k-1)$, $\la(2k)$ belongs to $\{1,\dots,n\}$ and the other
one belongs to $\{n+1,\dots,2n\}$. We get
\[ \frac{4\pi r^{2n}}{(2n+1)!}2^n n!\sum_{\rho\in \Pi_n}\sum_{i,j}Q^{i_1\dots i_n}S^{j_1\dots j_n} \de_{i_{\rho(1)}j_1}\dots \de_{i_{\rho(n)}j_n}, \]
which due to the symmetry of $Q$ gives the right hand side of
\rf{C.3}. For $m<n$ a similar arguments show that all terms vanish
(we use the traceless condition for $Q$ if $m+n$ is even and the
antisymmetry of the expression under the integral if $m+n$ is
odd). \ed

Let $n=0,1,2,\dots$, $\qb$ be a tracelless symmetric tensor of
rank $n$, $f$ be an even analytic function of $r$ near $0$ ($f$ is
a constant for $n=0$). We set
\[ D_{n\qb f}=f\frac{Qx\cdots x}{r^{2n+1}} \]
and define $\fc$ as the linear span of all $D_{n\qb f}$. The
elements $F\in\fc$ become distributions if for any test function
$\rho$ we set
\be <F,\rho>=\int_0^{\infty} dr\int_{S^2(r)} r^2F\rho d\si=\lim_{R\ra 0^+}\int_{\rb^3\setminus K(0,R)} F\rho d^3x. \et{C.4} \ee
That is well defined because setting $F=D_{n\qb f}$ and using
\be \rho = \sum_{k=0}^{n-1}\frac1{k!} \rho_{i_1\dots i_k} x^{i_1}\dots x^{i_k}
+ O(r^n),\et{C.5}\ee
\be \rho_{i_1\dots i_k}=(\p_{i_1}\dots\p_{i_k}\rho)(0), \et{C.5a} \ee
we obtain
\[ \int_{S^2(r)} r^2F\rho d\si=\left[\sum_{k=0}^{n-1}\frac1{k!}\frac{f}{r^{2n-1}}\int_{S^2(r)} (Qx\cdots x)(\rho x\cdots x)d\si\right]+O(r)=O(r), \]
since all the integrals over $S^2(r)$ vanish due to $k<n$ and
Proposition 2.

For $F\in\fc$ let $\p_i$ denote the partial derivative in the
sense of distributions, $\p_i^C$ - the partial derivative as
function,
\be \p_i^R F=\p_i F-\p_i^C F. \et{C.6} \ee
We define $\lb_i$ as tensor of rank $1$ such that
$(l_i)_j=\de_{ij}$.

{\bf Proposition 3.}
\[ x_iD_{n\qb f}=D_{n+1,\lb_i\vee\qb,r^2f}+\frac{n}{2n+1}D_{n-1,\lb_i\ip\qb,f}, \]
\[ \p_i^C D_{n\qb f}=D_{n+1,\lb_i\vee\qb,-(2n+1)f+rf,r}+\frac{n}{2n+1}D_{n-1,\lb_i\ip\qb,f_{,r}/r}. \]
{\em Proof.} We use $x_i(Qx\cdots x)=(\lb_i\vee\qb)(x\cdots
x)+\frac{n}{2n+1}r^2Q_i$, where  $Q_i=Q_{ii_2\dots
i_n}x^{i_2}\dots x^{i_n}=(\lb_i\ip\qb)(x\dots x)$. \ed

Thus for $F\in\fc$, $\p_i^CF$ and $\p_i^RF$ are distributions.

{\bf Proposition 4.} For $F\in\fc$ and a test function $\rho$
\[ <\p_i^RF,\rho> =\lim_{R\ra 0^+}\int_{S^2(R)}F\rho x_i R\, d\si. \]
{\em Proof.} Due to \rf{C.4}
\[ <\p_iF,\rho>=-<F,\p_i\rho>=\lim_{R\ra 0^+}\int_{\rb^3\setminus K(0,R)} (\p_i^CF)\rho d^3x - \lim_{R\ra 0^+}\int_{\rb^3\setminus K(0,R)} \p_i^C(F\rho)d^3x, \]
where the first term equals $<\p_i^C F,\rho>$. Setting
$(G_i)_j=F\rho\de_{ij}$ and using \rf{C.6}, one gets
\[ <\p_i^R F,\rho>= -\lim_{R\ra 0^+}\int_{\rb^3\setminus K(0,R)} \di G_i\, d^3x
\]
         \[ = \lim_{R\ra 0^+}\int_{S^2(R)}(G_i)_j\frac{x^j}R
R^2d\si=\lim_{R\ra 0^+}\int_{S^2(R)}F\rho x_iR\,d\si. \quad
\mbox{Q.E.D.} \]

{\bf Proposition 5.}
\[ \p_i^R D_{n\qb f}=[\lim_{r\ra 0} f(r)](-1)^{n-1}\frac{4\pi n}{(2n+1)!!}Q^{iI}\p_I\dt, \]
where $I=(i_2,\dots i_n)$, $\p_I=\p_{i_2}\dots\p_{i_n}$.

{\em Proof.} Using Proposition 4 for $F=D_{n\qb f}$, \rf{C.5},
setting $c=\lim_{r\ra 0}f(r)$, $(\rho_i)^{i_1\dots i_k
i_{k+1}}=\rho^{i_1\dots i_k}\de_i^{i_{k+1}}$ and using Proposition
2, we get
\[ \rho_i(x\cdots x)=\rho^{i_1\dots i_k}x_{i_1}\cdots
x_{i_k}x_i,\]
\[ <\p_i^R D_{n\qb f},\rho>=
\sum_{k=0}^{n-1}\frac1{k!}\lim_{R\ra
0^+}\int_{S^2(R)}f(R)\frac{\rho_i(x\cdots x)Q(x\cdots
x)}{R^{2n}}d\si=
\]
\[ \frac{c}{(n-1)!}Q^{ii_2\dots i_n}\rho_{i_2\dots i_n}\frac{4\pi(n!)^2 2^n}{(2n+1)!}. \]
Moreover, we use $<\p_I\dt,\rho>=(-1)^{n-1}\rho_{i_2\dots i_n}$
(cf. \rf{C.5a}) and $\frac{(2n+1)!}{n!2^n}=(2n+1)!!$ \ed

{\bf Proposition 6.} Assuming $f\in L_{n0}$ (cf. Section 4) and
setting $g=fr^n$ (which is regular at $0$), we obtain
\[ X_{k(n)}(f,\qb)=D_{n+1,\lb_k\vee\qb,(2n+1)g-rg_{,r}}+ \frac{n+1}{2n+1}D_{n-1,\lb_k\ip\qb,g_{,r}/r}, \]
\[ Y_{k(n)}(f,\qb)=-D_{n,\lb_k\times\qb,g/\ph}. \]
{\em Proof.} By a direct computation. \ed

{\bf Proposition 7.}
\[ \p_i^R X_{k(n)}(r^{-n},\qb)=(-1)^n\frac{4\pi(2n+1)}{(2n+3)!!}\times \]
\[ \times[\de_{ik}Q^J\p_J\dt+nQ_{i}{}^I\p_{kI}\dt-
\frac{n(n-1)}{2n+1}Q_{ik}{}^L\p^{m}{}_{mL}\dt
-\frac{2n}{2n+1}Q_{k}{}^I\p_{iI}\dt], \]
\[ \p_i^R X_{k(n)}(r^{-(n-2)},\qb)=(-1)^n\frac{4\pi(n-1)}{(2n-1)!!} Q_{ik}{}^L\p_{L}\dt, \]
\[ \p_i^R X_{k(n)}(f,\qb)=0 \mbox{ if } \lim_{r\ra 0}(fr^{n-2})=0, \]
\[ \p_i^R Y_{k(n)}(r^{-n},\qb)=(-1)^n\frac{4\pi}{(2n+1)!!} [(n-1)\eps_k{}^{ma}Q_{mi}{}^L\p_{aL}\dt +\eps_{ik}{}^m Q_m{}^I\p_{I}\dt], \]
\[ \p_i^R Y_{k(n)}(f,\qb)=0 \mbox{ if } \lim_{r\ra 0}(fr^n)=0. \]
{\em Proof.} We use Proposition 6 and Proposition 5. \ed

{\em Remark.} In the case of $n=0,1$ some equations related to the
first equation of Proposition 7 were presented in (6)--(7) of
\cite{Frahm}.

Next, we obtain that in addition to \rf{14b}, \rf{19}--\rf{20},
\rf{24}--\rf{25}, the operations $\di$, $\rotf$ and $\#$ acting on
the fields $\xb$ and $\yb$ yield the following distribution parts:

{\bf Proposition 8.}
\[ \di^R \xb_{(n)}(r^{-n},\qb)=(-1)^n\frac{4\pi(n+1)}{(2n+1)!!}Q^J\p_J\dt, \]
\[ [\rotf^R \xb_{(n)}(r^{-n},\qb)]_k= (-1)^n\frac{4\pi n}{(2n+1)!!}\eps_{ki}{}^jQ^{iI}\p_{jI}\dt,\]
\[ [\xb_{(n)}(r^{-n},\qb)^{\# R}]_k=(-1)^n\frac{4\pi n(n^2-1)}{(2n+1)!!}Q_{ki}{}^Lb^i\p_L\dt, \]
\[ \left[\rotf^R \yb_{(n)}(r^{-n},\qb)\right]_k =(-1)^n\frac{4\pi(n+1)}{(2n+1)!!}Q_k{}^I\p_I\dt, \]
\[ \di^R \yb_{(n)}(r^{-n},\qb)=\yb_{(n)}(r^{-n},\qb)^{\# R}=0, \]
while for $f$ such that $\lim_{r\ra 0}(fr^n)=0$ the distribution
parts vanish.

{\em Proof.} We use Proposition 7 and notice that the operator
$d/d\tau$ appearing in $\#$ (cf. \rf{12a}) gives no additional
terms. \ed

Let us notice that $\psi_0=D_{011}=\frac1r$ is the classical
monopole potential, $-\Delta\psi_0=4\pi\dt$. By induction one has
$Q^I\p_I\psi_0=(-1)^n(2n-1)!!D_{n\qb1}$ (no distribution parts! -
cf. Proposition 5) and therefore $\psi_n=D_{n\qb1}$ satisfies
\[ -\Delta\psi_n=(-1)^n\frac{4\pi}{(2n-1)!!}Q^I\p_I\dt \]
(cf. \rf{28}) and $\psi_n$ is the classical $2^n$--pole potential.
Using \rf{C.6} and Propositions 3,5 and 6, we obtain that the
corresponding electric field is given by
\be \dc_n^{cl}=-\grad\psi_n, \quad (\dc_n^{cl})_k=X_{k(n)}(r^{-n},\qb)+(\dc_d)_k, \et{C.7}\ee
where
\[ (\dc_d)_k=(-1)^n\frac{ 4\pi n}{(2n+1)!!}Q_k{}^I\p_I\dt \]
(cf. \rf{DD} and Remark 2 in Section 5). After some computations
one obtains
\be \di\dc_d=(-1)^n\frac{ 4\pi n}{(2n+1)!!}Q^J\p_J\dt, \et{C.8} \ee
\be (\rotf\dc_d)_k=(-1)^{n+1}\frac{4\pi n}{(2n+1)!!}\eps_{ki}{}^jQ^{iI}\p_{jI}\dt, \et{C.9}\ee
\be (\dc_d^{\#})_k=(-1)^n\frac{ 4\pi n}{(2n+1)!!}\dot Q_k{}^I\p_I\dt. \et{C.10} \ee

{\em Proof of Theorem 2.} Let us first omit the terms containing
$\dt$ and its derivatives. Then $\dn{l}$ and $\bn{l}$ consist of
$\xb$ and $\yb$ fields and therefore (cf. \rf{14b})
$\di\dn{l}=\di\bn{l}=0$, which proves \rf{9}--\rf{10}. Moreover,
applying $\rotf$ to \rf{33}--\rf{34}, we obtain
\be \rotf\bn{k+1}=\dn{k}\ha, \quad k=-n-2,-n,-n+2,\dots, \et{36}\ee
\be -\rotf\dn{k+1}=\bn{k}\ha, \quad k=-n-1,-n+1,-n+3,\dots. \et{37}\ee
But one has also
\be \rotf\dn{-n-2}=0 \et{38}\ee
(we use \rf{20} and $h_{(n)}[P_n(z)]=0$). Inserting
\rf{30}--\rf{31} into the both sides of \rf{11}--\rf{12} and using
\rf{36}--\rf{38}, one proves that \rf{11}--\rf{12} are fulfilled.

Let us notice that $\dt$ and its derivatives appear in the sources
\rf{28}--\rf{29}, in $\dc_d$ and also when acting by $\di$,
$\rotf$ or $\#$ on $\xb_{(k)}(r^{-k},{\bf P})$,
$\yb_{(k)}(r^{-k},{\bf P})$. Using \rf{22}--\rf{25}, one shows
that $\xb_{(k)}(r^{-k},{\bf P})$, $\yb_{(k)}(r^{-k},{\bf P})$
appear only in $\dn{-n-2}$ which contains $\xb_{(n)}(r^{-n},\qb)$
and in $\bn{-n-1}$ which contains
\[ -\frac{n^2-1}{2n+1}\yb_{(n-1)}(r^{-(n-1)},\bb\ip\qb)-\yb_{(n)}(r^{-n},\dot\qb).\]
 Therefore $\di\dc$ contains additionally
\[ \di\dc_d+\di^R \xb_{n}(r^{-n},\qb)=(-1)^n\frac{4\pi}{(2n-1)!!}Q^J\p_J\dt=\jc^0, \]
$\di\bc$ has no additional terms, $\dc\ha-\rotf\bc$ contains
additionally (cf. Proposition 8, \rf{C.10} and \rf{29})
\[ \dc_d\ha +\xb_{(n)}(r^{-n},\qb)^{\# R}+\frac{n^2-1}{2n+1}\rotf^R\yb_{(n-1)}(r^{-(n-1)},\bb\ip\qb) \]
\[ +\rotf^R\yb_{(n)}(r^{-n},\dot\qb)=-\jc, \]
$\bc\ha+\rotf\dc$ contains additionally $\rotf\dc_d+\rotf^R
\xb_{(n)}(r^{-n},\qb)=0$, hence the Maxwell equations are
satisfied also in the distributional sense. The last statement of
Theorem 2 follows from the properties of $\#$, $\rotfi$ and $\xb$
and $\yb$ fields.\ed

\end{appendix}

\begin{center} {\bf Acknowledgements} \end{center}

We thank Dr Marcin Ko\'scielecki for fruitful discussions.

\end{document}